\documentclass[conference]{IEEEtran}
\ifCLASSINFOpdf
  \usepackage[pdftex]{graphicx}
\else
  \usepackage[dvips]{graphicx}
\fi

\usepackage[cmex10]{amsmath}
\usepackage{amssymb}
\def\RTN{\scriptscriptstyle RTN}
\begin{document}

\title{Time-evolution of entanglement and quantum discord of 
bipartite systems subject to 1/$f^\alpha$ noise}

\author{\IEEEauthorblockN{Claudia Benedetti and Matteo G. A. Paris}
\IEEEauthorblockA{Dipartimento di Fisica \\ Universit\`a degli Studi  di Milano \\ I-20133 Milano, Italy\\
Email: {claudia.benedetti@unimi.it}\\
Email: {matteo.paris@fisica.unimi.it}}
\and
\IEEEauthorblockN{Fabrizio Buscemi}
\IEEEauthorblockA{ARCES\\Universit\`a di Bologna\\
I-40125 Bologna, Italy\\
Email: {fabrizio.buscemi@unimore.it}}
\and
\IEEEauthorblockN{Paolo Bordone}
\IEEEauthorblockA{Dipartimento di Scienze Fisiche,\\ Informatiche e Matematiche\\
Universit\`a di Modena e Reggio Emilia\\
I-41125 Modena, Italy\\
Email: {bordone@unimore.it}}}
\maketitle

\begin{abstract}
We study the dynamics of quantum correlations for two non interacting
qubits initially prepared in a maximally entangled state and then
coupled with an external environment characterized by a noise spectrum
of the form $1/f^\alpha$. The noise spectrum is due to the interaction
of each qubit with a collection of $N_f$ classical fluctuators with
fixed switching rates.  We find that,
depending on the characteristic of the noise spectrum considered, both
entanglement and quantum discord display either a monotonic decay or the
phenomena of sudden death and revivals.
\end{abstract}
\IEEEpeerreviewmaketitle

\section{Introduction}
Quantum correlations, entanglement and quantum discord (QD), represent 
a valuable resource for information processing and technology \cite{revent,revdisc}. 
In
turn, the dynamics of quantum correlations have
been investigated for several systems, ranging from quantum
optics~\cite{man07,vasile09,vasile11,vasile10,haikka12,genoni08,brida10}
to nanophysics~\cite{bus11,bus12,maziero09}.  
The major limitation of the use of entanglement and QD in practical 
applications is due to the unavoidable interaction of real quantum 
systems with their surroundings, resulting in decoherence processes, and, 
as a consequence, in the degradation of quantum correlations. 
On the other hand, the environment can even resume or preserve quantum 
correlations, as it happens when the backflow of information to the 
system becomes relevant. It is therefore of interest to understand the effect 
of the various kinds of environmental noise on the dynamics of quantum
correlations in realistic quantum systems, where peculiar phenomena can be 
observed, such as sudden death and
revival~\cite{yu04,yu09}. In particular, it is crucial to investigate and
compare Markovian noise, ascribed to environment with short
self-correlations~\cite{yu04,yu09}, to non-Markovian noise~\cite{bellomo07},
which is associated to environments with memory and may lead to a non-monotonic
time dependence of entanglement and QD.  Revival phenomena have been
predicted both for couple of qubits interacting directly or indirectly in
a common quantum reservoir~\cite{ficek06,mazzola09} and for
non-interacting qubits in independent non-Markovian quantum
environments.\cite{bellomo07}.
\par
The effect of the quantum noise on the entanglement dynamics between two
quantum systems has been interpreted in terms of the transfer of the
correlations back and forth from the systems themselves to the
environment. This is due to the back-action of the system on the
environment. On the other hand, recently, revivals of quantum
correlations have been found also for quantum systems coupled to
classical sources~\cite{zhou10} and have been connected to a
quantifier of non-Markovianity for the dynamics of a single-qubit.  It
was actually proven that a classical noise can mimic a quantum
environment not affected by the system or influenced in a way that does
not result in a back-action.
\par
In this work we analyze the role played by classical environments,
characterized  by $1/f^\alpha$ noise spectra, on the dynamics of quantum
correlations. In particular, as prototypical
bipartite system, we consider two non-interacting qubits coupled to
noise in different environments. Noises of the type $1/f^\alpha$ are
among the main sources of decoherence in quantum solid-state
devices~\cite{weissman88,kaku07,tsai06,paladino02,burkard,bellomo10,paladino11},
thus constituting a fundamental case of study for a deeper understanding
of the decoherence process itself. This kind of noise spectra stem from
a collection of random telegraph sources with a specific distribution of
their switching rates. For $\alpha=1$ the so-called pink $1/f$ noise is
found, which is obtained from a set of random telegraph fluctuators
weighted by the inverse of the switching rate. Another interesting case
is the $1/f^2$ spectrum, also called brown noise from its relation to a
Brownian motion.  We study the dynamics of quantum correlations as a
function of the parameter $\alpha$ and observe a continuous variation in
their bahavior, ranging from the monotonic decay characteristic of the
$1/f$ noise to the phenomena of sadden death and revivals induced by the
$1/f^2$ noise. While a qualitative trend is clearly identified, it is
not possible to identify a precise threshold value between the decaying
and the oscillating regime, since the outcomes significantly depend upon
the range of selected frequencies and upon the number of random
telegraph fluctuators used for the calculations.  The dynamics of the
two quits is ruled by a stochastic Hamiltonian with time dependent
coupling. The average of the time-evolved states over the switching
parameters describes the evolution of the two-qubit state under the
effect of the noise.
\par
The paper is organized as follows: in Sec. II the definition of
negativity, as a measure of entanglement, and of QD are briefly
reviewed.  The physical model adopted is described in Sec. III. In Sec.
IV results are presented and discussed, whereas Section \ref{s:out}
closes the paper with some concluding remarks.
\section{Measures of quantum correlations}
Entanglement is evaluated by means of negativity, defined as:
$N=2\left|\sum_i\lambda_i^{-}\right|$
where $\lambda^-_i$ are the negative eigenvalues of the partial
transpose of the density matrix of the system. Negativity is zero 
for separable states and assumes value one for maximally entangled 
states. 
\par
QD is defined as the difference between the total and the classical
correlations in a system 
$\mathcal{Q}=\mathcal{I}-\mathcal{C}$,
where $\mathcal{I}$ is  the quantum mutual information 
$\mathcal{I}=S(\rho^A) +S(\rho^B)-S(\rho)$, 
quantifying the total correlations in a system, $S$ is the von Neumann
entropy, and $\rho^{A(B)}$ indicates
the reduced density matrix of the subsystem $A(B)$, 
$\mathcal{C}$ denotes  the measurement-induced quantum mutual information,
namely the classical correlations, and reads
$\mathcal{C}=\max_{\{\Pi_k\}}\{ S(\rho_A)-S(\rho|\{\Pi_k\})\}$, 
with $S(\rho|\{\Pi_k\})$ that denotes the quantum conditional entropy 
with respect to the set of projective measurements $\{\Pi_k\}$ performed 
locally  on subsystem $B$. Usually to compute QD is not an easy task, since 
it involves a maximization procedure.  However, for two-qubit systems described
by a density matrix with a so called ``X'' form,  an analytical expression for 
$\mathcal{Q}$ has been obtained \cite{luo08}.
\section{Kinematics and dynamics}
We consider a system of two non-interacting qubits, initially entangled,
subject to a noisy classical environment with noise spectra of the type
$1/f^\alpha$, with $1\le\alpha\le 2$. The environment is modelled by
using different configurations of bistable fluctuators. The two qubits
are initially prepared in the Bell state
$|\phi^+\rangle=(|00\rangle+|11\rangle)/\sqrt{2}$.  Here the interaction
between the qubits and the environment is assumed local, i.e. the two
qubits interact with two different and independent baths.  The case of a
common environment affecting both qubits and its comparison with the
local model is discussed in~\cite{benedetti12}.  Setting $\hbar=1$ and
adopting the spin notation, the two-qubit Hamiltonian describing the
interaction with a single fluctuator is given by
$H(t)=H_A(t)\otimes\mathbb{I}_B+\mathbb{I}_A\otimes H_B(t)$,
where $H_{A(B)}$ is the Hamiltonian of a single qubit
subject to a classical time-dependent noise which affects the transition
amplitude parameter $c_{A(B)}(t)$: \begin{equation}\label{h1qubit}
H_{A(B)}(t)=\epsilon \mathbb{I}_{A(B)}+\nu c_{A(B)}(t)\sigma_{x\;A(B)},
\end{equation} with $\epsilon$ the qubit energy in absence of noise
(energy degeneracy is assumed), $\mathbb{I}_{A(B)}$  the identity matrix
for subspace $A(B)$, $\nu$ is the coupling constant between the system
and the environment, $\sigma_x$ the Pauli matrix.  If the time-dependent
coefficient $c_{A(B)}(t)$ can randomly flip between two values
$c(t)=\pm1$ with a fixed rate $\gamma$, then Eq. \eqref{h1qubit}
describes a qubit subject to a random telegraph noise (RTN)
\cite{benedetti12,benedetti13,bergli09,mazzola11,bordone12}. This model 
has recently been extended
to the case of tripartite systems~\cite{buscemi13}. 
The above Hamiltonian is stochastic due to
the random nature of the noise parameter $c(t)$. For a specific choice
of $c(t)$, the total system evolves  according to the evolution operator
$e^{-i\int H(t')\:dt'}$.  By averaging the global state over different
realizations of the sequences of  $c(t)$, the two-qubit mixed state is
obtained. 
\par
In order to reproduce the $1/f^{\alpha}$ spectrum, the
single RTN frequency power density must be integrated over the switching
rates $\gamma$ with a proper distribution:
\begin{align}
 S_{1/f^{\alpha}}(f)=\int_{\gamma_1}^{\gamma_2}S_{\RTN}(f,\gamma)\ p_{\alpha}(\gamma)\,d\gamma\label{spectrum},
\end{align}
where $S_{\RTN}(f,\gamma)$ is the random telegraph noise frequency
spectral density with Lorentzian form $S_{\RTN}(f,\gamma)=4\gamma/(4\pi^2
f^2+\gamma^2)$.  The integration is performed between a minimum and a
maximum value of the switching rates, respectively $\gamma_1$ and
$\gamma_2$.  $p(\gamma)$ is the switching rate distribution and takes a
different form depending on the kind of noise:
\begin{align}\label{distrib}
p_{\alpha}(\gamma)=\left\{\begin{array}{ll}
 \frac{1}{\gamma\,\ln(\gamma_2/\gamma_1)}&\alpha=1\\
 & \\
\frac{(\alpha-1)}{\gamma^{\alpha}}
 \left[\frac{(\gamma_1\gamma_2)^{\alpha-1}}
 {\gamma_2^{\alpha-1}-\gamma_1^{\alpha-1}}\right]&1<\alpha\leqslant2
\end{array}
\right.\ .
\end{align}
It follows that, in order to simulate a frequency spectrum proportional
to $1/f^{\alpha}$, the switching rates must be selected from a
distribution proportional to $1/\gamma^{\alpha}$.  When  the integration in Eq. 
\eqref{spectrum} is performed, the spectrum has the requested
$1/f^{\alpha}$ behavior in a frequency interval, so that every frequency belonging to such interval satisfies the condition
 $\gamma_1\ll f\ll\gamma_2$. Eq. \eqref{spectrum} is obtained using a collection of sources of RTN each with
 a switching rate taken from the same $\gamma$-distribution.
   
The $1/f^{\alpha}$ noise spectrum is obtained from the coupling of the system with a large number of fluctuators, each characterized by a specific switching rate, picked from the distribution $p_\alpha(\gamma)$~\cite{bellomo10} in a range $[\gamma_1,\gamma_2]$.
In this case the random parameters in Eq.\eqref{h1qubit} describes a linear combination
of bistable fluctuators $c(t)=\sum_{j=1}^{N_f}c_j(t)$, where $N_f$ is
the number of fluctuators and we drop the subscript $A(B)$ to simplify
the notation.  Each $c_j(t)$ has a lorentzian power spectrum whose sum
gives the power spectrum of the noise:
\begin{align}
S(f)=\sum _{j=1}^{N_f}S_j(f;\gamma_j)=\sum_{j=1}^{N_f}\frac{\gamma_j}{\gamma_j^2+4\pi^2f}\propto \frac{1}{f^{\alpha}}\ .
\end{align}
In order to obtain a reliable $1/f^\alpha$ spectrum, it is necessary that a sufficiently large number of fluctuators is considered, and
that the selected $\gamma_j$ are a representative sample of the distribution $p_\alpha(\gamma_j)$ 
in the range $[\gamma_1,\gamma_2]$. We assume that all the fluctuators have the same coupling constant with the environments,
that is $\nu_j=\nu$ for $j=1,...,N_f$.
\par
For a single fluctuator, let us say the $j^{th}$ one, the
global system evolves according to the Hamiltonian $H(t)$
with a specific choice of both the parameter $c_j(t)$ and of its 
switching rate $\gamma_j$. The interaction with a bistable fluctuator 
induces a phase shift in the state of each single qubit given by
$
\varphi_j(t)=-\nu\int_0^tc_j(t')dt'
$
and characterized by a distribution \cite{bellomo10,bergli09}:
\begin{align}\label{distribution_phi}
 &p(\varphi_j,t)= \frac{1}{2}e^{-\gamma_j t}\nonumber \\ 
 &\times\left\{[\delta(\varphi_j+\nu t)+\delta(\varphi_j-\nu t)] +
 \frac{\gamma_j}{\nu}[\Theta(\varphi_j+\nu t)+\Theta(\varphi_j-\nu t)]\right.\nonumber\\
&\left.\times\left[\frac{I_1\left(\gamma_j t\sqrt{1-(\varphi_j/\nu t)^2}\right)}{\sqrt{1-(\varphi_j/\nu t)^2}}
+I_0\left(\gamma_j t\sqrt{1-(\varphi_j/\nu t)^2}\right)\right]\right\}
\end{align} 
where $I_v(x)$ is the modified Bessel function and 
$\Theta(x)$ is the Heaviside step function.
For a given $\gamma_j$ the global system is described by a X-shaped
density matrix obtained by averaging over the noise the density
matrix $\rho(\varphi_j,\gamma_j,t)$ corresponding to a specific choice of
the parameter $c_j(t)$ \cite{benedetti12} : 
\begin{align}
\rho(\gamma_j,t)&=\int \rho(\varphi_j,\gamma_j,t) p(\varphi_j,t)d\varphi_j\nonumber\\
&=\frac{1}{2}\Big[(1+D^2_j(t))|\phi^+\rangle\langle\phi^+|
+(1-D^2_j(t))|\psi^+\rangle\langle\psi^+|\Big]\label{rtn}
\end{align}
where  $|\psi^+\rangle=(|01\rangle+|10\rangle)/\sqrt{2}$ is a Bell state.
The function $D_j(t)$ represents the average of the RTN phase factor i.e. 
$
\langle e^{2i\varphi_j(t)}\rangle=\int e^{2i\varphi_j(t)}p(\varphi_j,t)\,d\varphi_j=D_j(t)
$
where
\begin{equation}
\label{ddtdif}
D_j(t)=\left\{ 
\begin{array}{l}e^{-\gamma_j t}\left[\cosh {\left(\kappa_\nu t\right)} 
+\frac{\gamma_j}{\kappa_\nu}\sinh {\left(\kappa_\nu t\right)}\right]\hbox{for}\ \gamma_j \geq 2\nu
 \\ 
 e^{-\gamma_j t}\left[\cos {\left(\kappa_\nu t\right)} +
 \frac{\gamma_j}{\kappa_\nu}\sin {\left(\kappa_\nu t\right)}\right]\hbox{for}\ \gamma_j \leq 2\nu
\end{array}\right . ,
  \end{equation}
with $\kappa_\nu=\sqrt{|\gamma_j^2 -(2\nu)^2|}$.
\par
In case of a collection of fluctuators, $j=1,...,N_f$, the global
evolution operator $U(t)\propto e^{-i\sum_j\varphi_j(t)}$, for fixed
values of the parameters associated to each fluctuator, permits to
compute the density matrix of the global system as a function of a total
noise phase $\varphi(t)=\sum_j\varphi_j(t)$.  Following the approach
used in \cite{benedetti12} to evaluate the dynamics of two qubits
subject to a single RTN, the time-evolved density matrix of the system
at time $t$ can be expressed as: 
$\overline{\rho(t)}=\int \rho(\varphi,t) p_T(\varphi,t)d\varphi$ 
where $p_T(\varphi,t)=\prod_j p(\varphi_j,t)$ is the global
noise phase distribution. This form of the density matrix depends
on the average of the phase factor $e^{2i\varphi(t)}$, which can be
computed in terms of the RTN coefficient $D(t)$ of Eq.\eqref{ddtdif}. We have
$
\langle e^{2i\varphi(t)}\rangle=\langle \prod_j e^{2i\varphi_j}\rangle=\prod_j D_j(t),
$
where the last equality holds since the RTN phase coefficients are independent.
Upon inserting the above expression in $\overline{\rho(t)}$ one may evaluate the two-qubit 
density matrix:
\begin{equation}\label{rho_many_fl}
\overline{\rho}(t, \, \{\gamma_j\})=
\frac{1}{2}\Big[(1+\Gamma)|\phi^+\rangle\langle\phi^+|
+(1-\Gamma)|\psi^+\rangle\langle\psi^+|\Big]\ ,
\end{equation}
where
$
\Gamma=\prod_j D_{jA}(t)D_{jB}(t)$,
where the subscripts $A$ and $B$ are reintroduced to underline that the result is derived from the contributions of the two subsystems.
Note that the fluctuators have fixed switching rates $ \{\gamma_j\}$
, $j=1\dots N_f$.
\begin{figure}[!t]
\centering
\includegraphics[width=0.4\columnwidth]{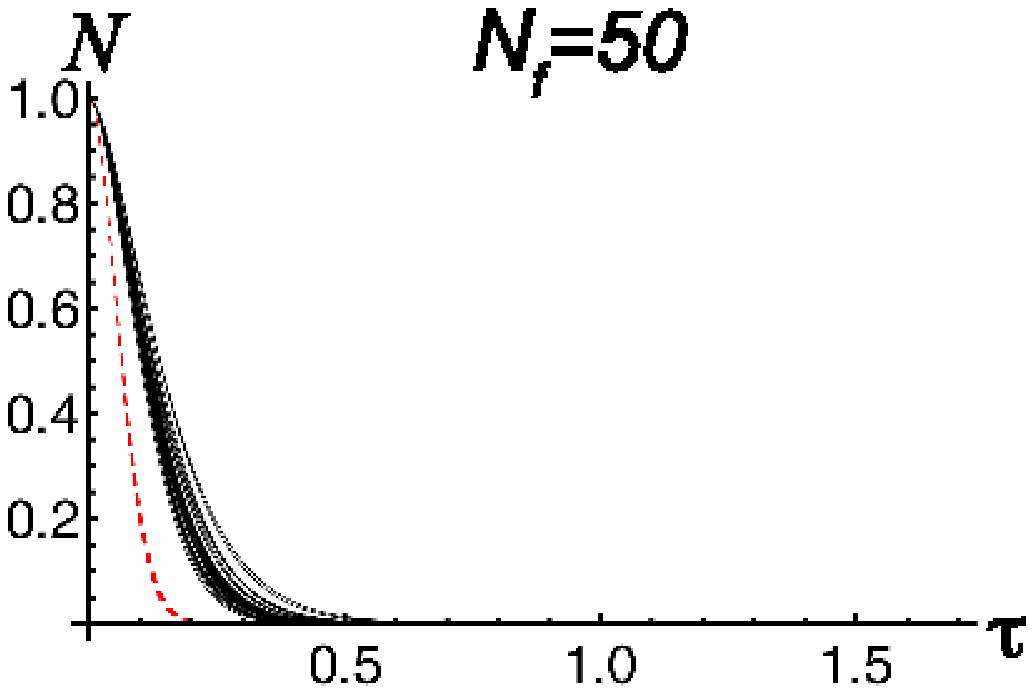}
\includegraphics[width=0.4\columnwidth]{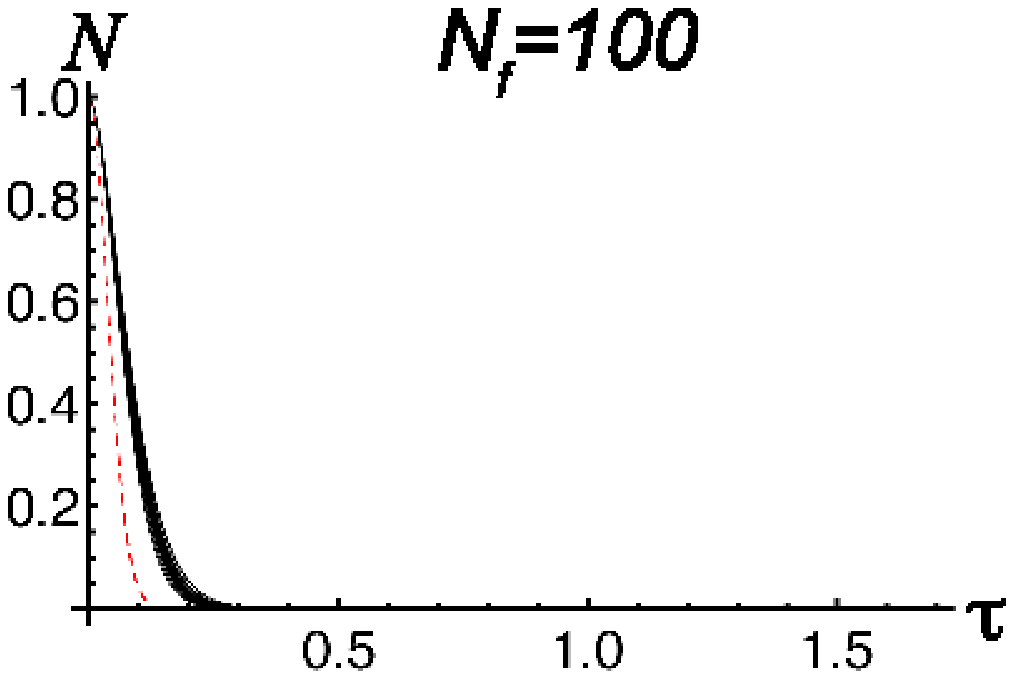}
\includegraphics[width=0.4\columnwidth]{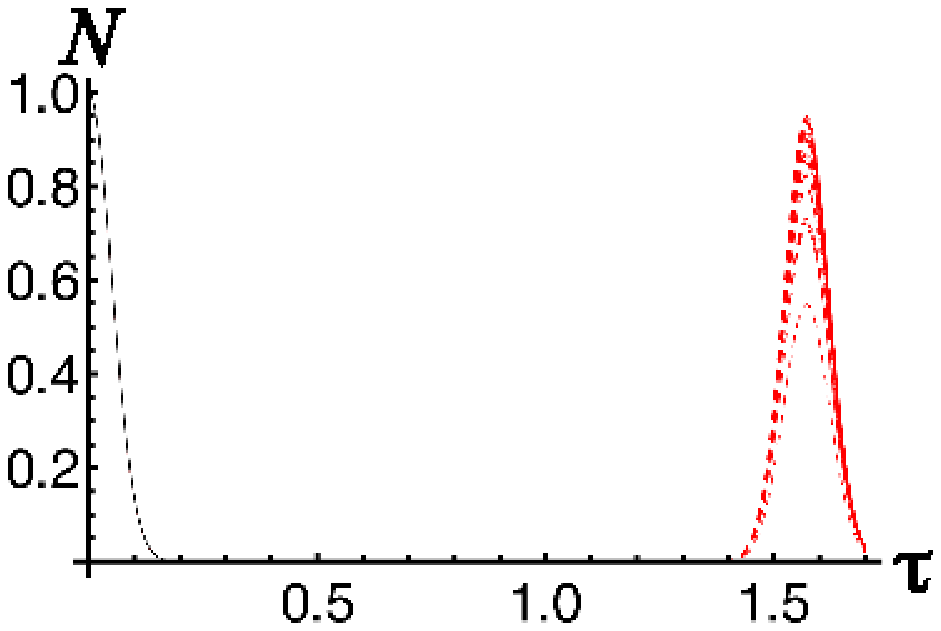}
\includegraphics[width=0.4\columnwidth]{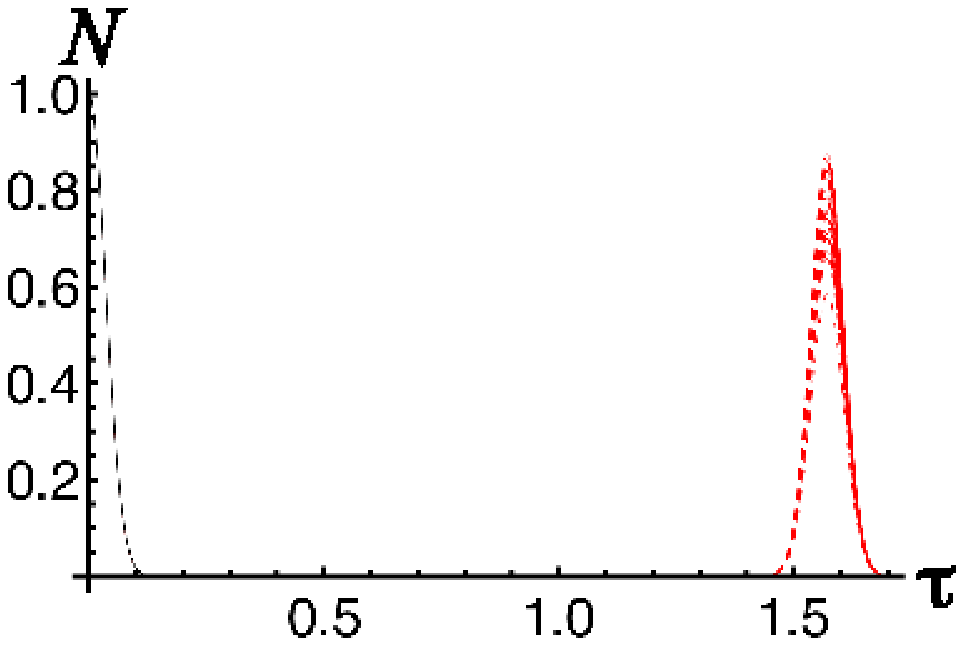}
\includegraphics[width=0.4\columnwidth]{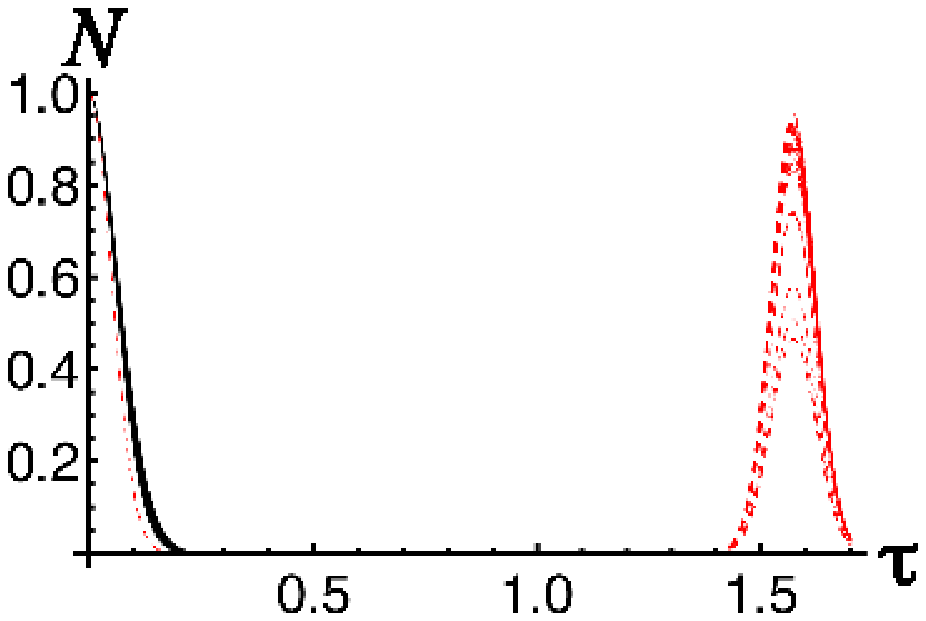}
\includegraphics[width=0.4\columnwidth]{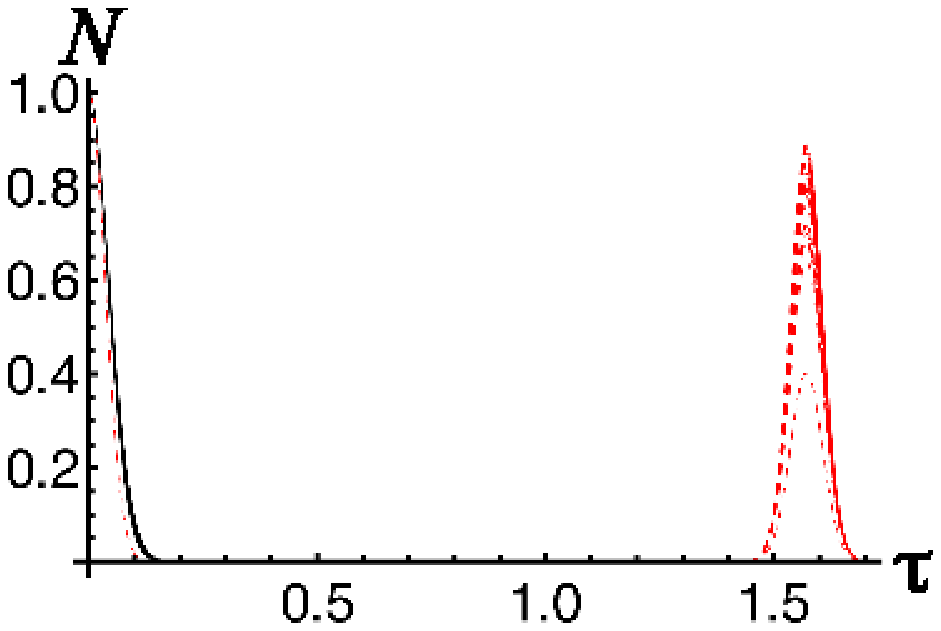}
\caption{(Color on line) Time evolution of the negativity for two numbers of bistable fluctuators $N_f=50$ (left column) and $N_f=100$
 (right column), and for two specific values of the noise parameter $\alpha$: $\alpha=1$ solid-black curves, $\alpha=2$ deshed-red curves. 
 30 curves are shown for each value of $\alpha$, corresponding to 30 different samples of the $\gamma_j$'s. 
 Different choices of the range $[\gamma_1,\gamma_2]/\nu$ are presented. Specifically:
$[1, 10^4]$ top panels, $[10^{-4}, 1]$ centre panels, $[10^{-4}, 10^4]$ bottom panels. 
Note that $\tau$ is a dimensionless parameter proportional to time.}
\label{fig1}
\end{figure}
\begin{figure}[!t]
\centering
\includegraphics[width=0.4\columnwidth]{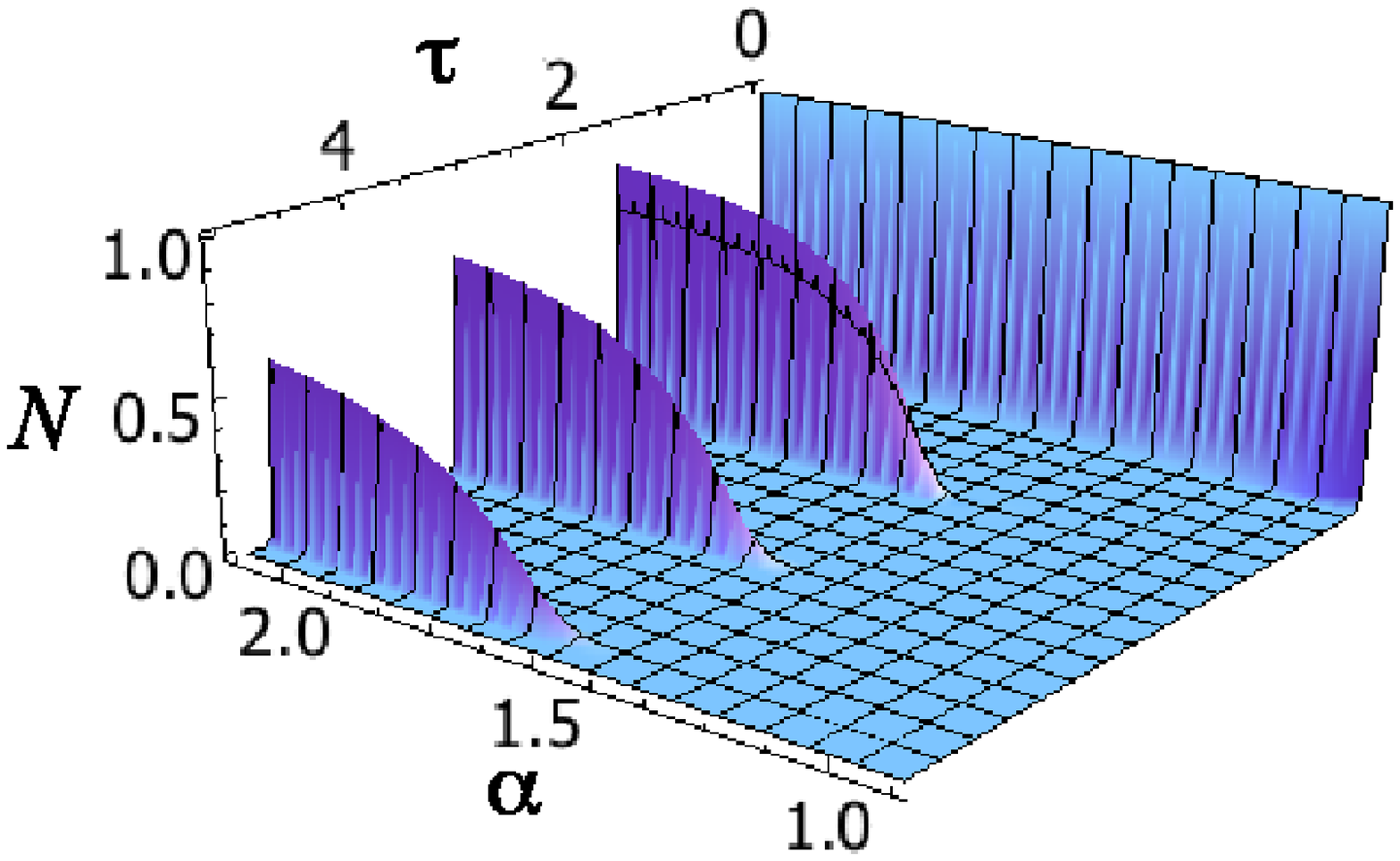}
\includegraphics[width=0.4\columnwidth]{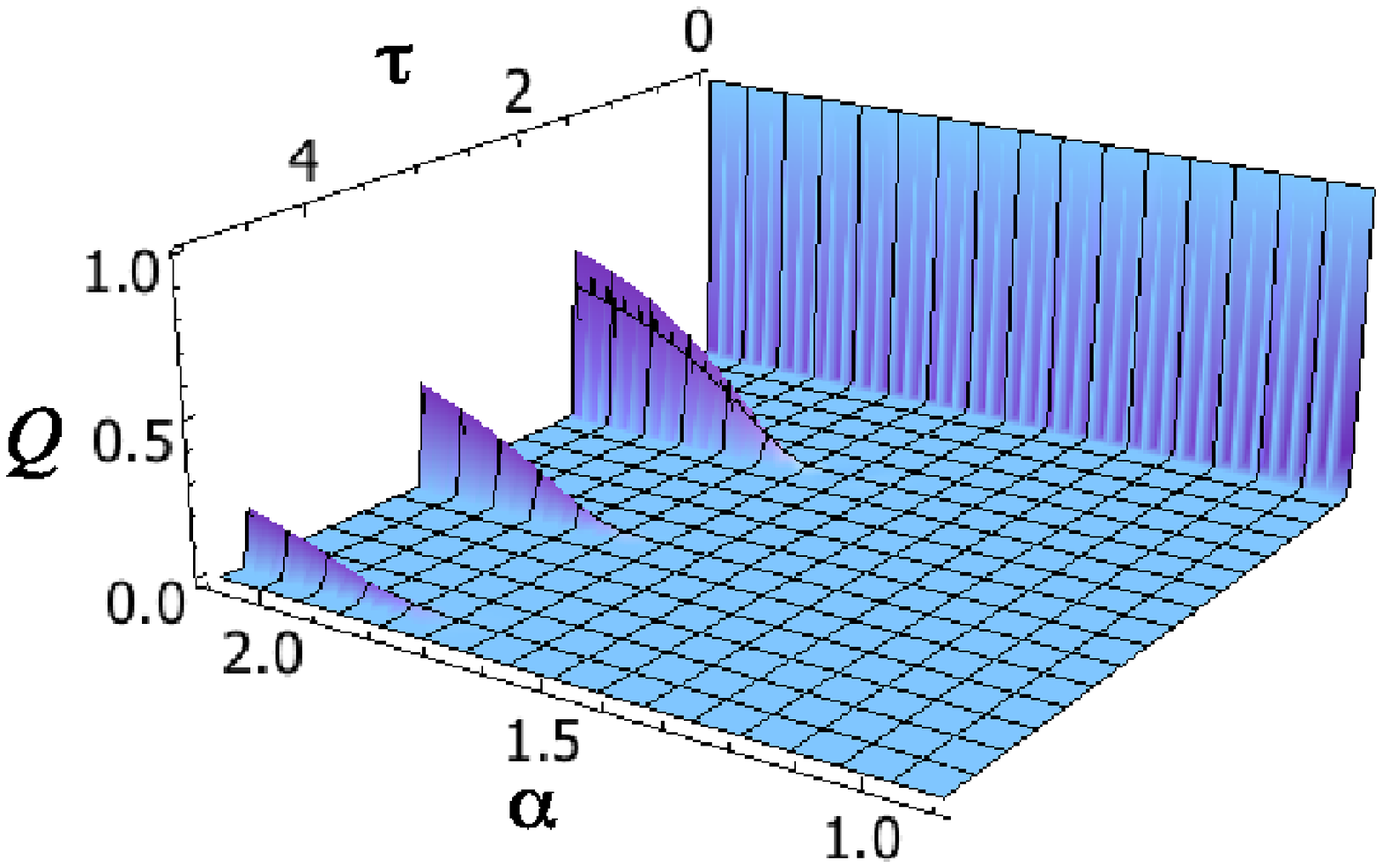}
\caption{(Color on line) Negativity (left panel) 
and quantum discord (right panel) as a function of the noise
parameter $\alpha$ and of dimensionless time $\tau$, for a 
specific sample of the $\gamma_j$'s and for $N_f=100$.} 
\label{fig2}
\end{figure}
\section{Results}
The density matrix in Eq. \eqref{rho_many_fl} preserves its X 
shape during time evolution. The negativity and the QD are thus
given by
\begin{equation}\label{QCS}
N(t)=\vert\Gamma(t)\vert \qquad Q(t)= h(\Gamma)\,,
\end{equation}
where
$ h(x) = \frac12[(1+x)\log_2{\left(1+x\right)} + 
(1-x)\log_2{\left(1-x\right)}]$.
Note that the choice of a maximally entangled initial state together
with the sheer dephasing nature of the interaction with the environment,
makes QD a function of negativity only. In these systems, in fact, the
evolved state is a mixture of Bell states.
\par
In Fig. \ref{fig1} we compare the time evolution of negativity for the
case of 50 and 100 fluctuators, and for two values of the parameter
$\alpha$, i.e.: $\alpha=1$ pink noise (solid-black line) and $\alpha=2$
brown noise (dashed-red line).  30 curves are drawn for each value of
$\alpha$, corresponding to 30 different samples of the $\gamma_j$'s.
Different choices of the range $[\gamma_1,\gamma_2]$ are confronted (see
figure caption). Increasing the number of fluctuators makes the results
less sensitive to the specific selection of the switching rates and
induces a faster decay of the quantum correlations, still the general
behavior is clearly maintained. For high frequencies both $1/f$ and
$1/f^2$ noise spectra induce an exponential decay of correlations, while
at low frequencies a revival peak appears for the $1/f^2$ spectrum,
whose height decreases increasing the number of fluctuators. Indeed a
number of peaks with periodicity $\pi/2$ and exponentially decreasing
heights are present. In Fig.\ref{fig1}, to make the picture more
readable, only the first one is shown. The periodicity can be understood
by analyzing the analytical expression of quantum correlations. 
The $D_j(t)$ functions
present damped oscillations for $\gamma<2\nu$ with periodicity
$2\pi/\kappa_\nu$, and for $\gamma>2\nu$ monotonically decay. Their
weighted superposition leads to an oscillating function with periodicity
$\pi/2$. Furthermore, for the $1/f^2$ noise spectrum, the selected
values of the $\gamma_j$'s accumulate near the lower value of the
frequency range, thus leading to a beat phenomenon with constructive
interference corresponding to the above mentioned periodicity with
narrow peaks.
\par
Fig.~\ref{fig2} reports negativity and QD as a function of time and of
the noise parameter $\alpha$, for a specific sample of the $\gamma_j$'s
and for $N_f=100$. The same qualitative behavior is found for such
quantities, due to their peculiar relation as given in (\ref{QCS}).
After the initial and fast exponential decay, at increasing values of
$\alpha$ the revival peaks, with $\pi/2$ periodicity, begin to appear.
The height of the peaks raises with $\alpha$ and reaches its maximum for
the $1/f^2$ noise spectrum.The qualitative trend is clearly identified,
while it is not possible to make precise claims about the heights of the
peaks or about a threshold value, for $\alpha$, regarding the appearance
of the peaks themselves. In fact, the quantitative results depends upon
the range of selected frequencies, upon the number of fluctuators, and
upon the $\gamma_j$'s sample. 
\section{Conclusions}\label{s:out}
In conclusions, we have investigated the effects of an external environment
characterised by a noise spectrum of the form $1/f^\alpha$ on the
dynamics of quantum correlations between two non interacting
qubits initially prepared in a maximally entangled state. 
The environment is modelled by an ensemble of bistable classical 
fluctuators, with fixed switching rates chosen from a given 
interval and a given distribution.
Our results
show that, depending on the characteristic of the noise spectrum,
quantum correlations display either a monotonic decay or the appearance
of revivals. The qualitative trend is clearly identified,
whereas the quantification of the results, and the identification of  the
threshold, depends explicitly on the range of selected frequencies, the 
number of fluctuators, and on the specific sample of the switching rates
$\gamma_j$.

\section*{Acknowledgment}
This work has been supported by MIUR (FIRB LiCHIS-RBFR10YQ3H) and
by a visiting scholarship of the Scottish Alliance of Physics Universities
(SUPA).

\end{document}